\documentclass[aps,prb,reprint,unsortedaddress,superscriptaddress]{revtex4-1}

\usepackage{amsmath,amssymb}

\usepackage{mediabb}
\usepackage{graphicx}

\usepackage{bm}
\usepackage{color}
\begin{document}

\title{Spin Berry phase in the Fermi arc states}

\begin{abstract}
Unusual electronic property of a Weyl semi-metallic nanowire is revealed.
Its band dispersion exhibits multiple subbands of partially flat dispersion,
originating from the Fermi arc states.
Remarkably,
the lowest energy flat subbands bear a finite size energy gap,
implying that electrons in the Fermi arc surface states
are susceptible of the spin Berry phase.
This is shown to be a consequence of spin-to-surface locking in the surface electronic states.
We verify this behavior and the existence of spin Berry phase 
in the low-energy effective theory of Fermi arc surface states 
on a cylindrical nanowire
by deriving the latter from a bulk Weyl Hamiltonian.
We point out that in any surface state exhibiting a spin Berry phase $\pi$,
a zero-energy bound state is formed 
along a magnetic flux tube of strength $\Phi_0/2 = hc /(2e)$.
This effect is highlighted in a surfaceless bulk system
pierced by a dislocation line, which shows
a 1D chiral mode along the dislocation line.

\end{abstract}

\date{\today}

\author{Ken-Ichiro Imura}
\affiliation{Department of Quantum Matter, AdSM, Hiroshima University, Higashi-Hiroshima 739-8530, Japan}

\author{Yositake Takane}
\affiliation{Department of Quantum Matter, AdSM, Hiroshima University, Higashi-Hiroshima 739-8530, Japan}

\maketitle

\section{Introduction}

It has been proposed recently that
a three-dimensional (3D) Weyl semimetal phase is 
likely realized in pyrochlore iridates,
A$_2$Ir$_2$O$_7$ (A=Y, or a lanthanide element).
\cite{Ashvin}
The Weyl semimetal has a unique band structure
characterized by a set of discrete linearly-dispersive point nodes, 
the Weyl points.
\cite{Balents_kiss, Volovik_1, Murakami_NJP}
Such a band structure of Weyl semimetal is naturally reminiscent of 
that of graphene, its 2D counter part, but unlike 2D Dirac cones:
$H_{2D} = p_x \sigma_x + p_y \sigma_y$,
3D Weyl points:
$H_{3D} = p_x \sigma_x + p_y \sigma_y + p_z \sigma_z$
are more stable objects;
they cannot be trivially gapped out by a perturbation
({\it cf.} $H_{2D}$ is gapped by a local mass term,
$H' = m\sigma_z$).

Unlike topological insulators also, the Weyl semimetal is gapless in the bulk,
but when in contact with the vacuum it exhibits
a peculiar surface state, somewhat reminiscent of a more established helical surface
state of the topological insulator.
The two surface states are indeed both of topological origin, and in this sense
this analogy between the Weyl semimetal and the topological insulator
is not superficial at all. 
In the bulk (without a surface) the Weyl semimetal is already gapless, 
but the conduction and the valence bands touch only at discrete points
(Weyl points) in the Brillouin zone -- let us assume hereafter that there exists only 
a pair of such point nodes for simplicity.
In the presence of a surface, an additional state appears, 
localized on the surface, and ``enveloping'' the two point nodes.
If one considers the $E=0$ cross section of the energy spectrum,
this additional surface state appears as a line, not necessarily straight
but always connecting the two Weyl points, forms a Fermi arc.
\cite{Ashvin, Balents_kiss, YBKim, Weyl_semi_2, Ran, Balents}
As is typically the case with the helical Dirac cone surface state
of a topological insulator, the existence of this Fermi arc envelope state
is {\it topologically protected} by a bulk topological invariant
through the so-called bulk/surface correspondence.

As we mentioned earlier, the Weyl semimetal can be regarded as
a 3D version of graphene.
Such an analogy in the low-energy electronic property of the {\it bulk} 
is naturally extended to that of the {\it surface}.
The Fermi arc state is indeed shown to be a precise 3D analogue of the edge states of 
a graphene nano-ribbon in the zigzag edge geometry.
\cite{Fujita}
These two examples constitute prototypical classes of the topologically non-trivial 
{\it gapless} states,
which are counterparts of the {\it gapped}
topological insulator and superconductors,
the latter known to be classified into the form of a periodic table (ten-fold way)
in terms of their symmetry and dimension.
\cite{Ryu_PRB, Ryu_NJP}
The idea of characterizing topologically non-trivial gapless or nodal states 
in terms of the topological invariants
has been introduced and extensively used in the study of $^3$He-A.
\cite{Volovik_1, Volovik_2, Volovik_3, Tsutsumi}
More recently it has been applied
to the description of topologically non-trivial nodal superconductors.
\cite{Sato_1, Sato_2, Yada_1, Yada_2, Schnyder_1, Schnyder_2, Schnyder_3}
A periodic table analogous to the one used for classifying various classes of
topological insulators and superconductors has been also proposed
for those classes of topologically non-trivial gapless states.
\footnote{S. Ryu, in ``Workshop and School on Topological Aspects on Condensed Matter Physics'', ICTP, Trieste, 27 June - 8 July; A. Schnyder, in ``Fifth Stig Lundqvist Conference on the Advancing Frontiers of Condensed Matter Physics'', ICTP, Trieste, 11-15 July, 2011.}

The edge/surface state of a topological insulator is often referred
to be ``helical'', indicating that its spin direction is locked with respect
to its propagating direction.
Here, in this work we focus on still another unique property of such a helical 
surface state, i.e., the existence of the spin Berry phase. 
The electronic spin in the surface helical state shows also spin-to-surface locking, 
\cite{spin_Berry_1, spin_Berry_2, Mirlin, JMoore, topo_weak, k2}
i.e., the spin is locked in-plane to the tangential surface of the real space geometry 
(e.g., on a cylindrical surface). 
A mathematical description of this spin-to-surface locking, the spin Berry phase 
primarily manifests in the finite-size energy gap associated with the surface helical states.
\cite{k2}
In an infinitely large (or doubly periodic) slab geometry, the finite-size energy gap 
of the surface state due to a finite thickness of the slab decays exponentially 
as a function of the thickness. 
In the case of a rectangular/cylindrical nanowire,
i.e., when the width of the slab becomes finite, 
and the slab acquires side surfaces, this is no longer the case. 
The phase information of the electronic wave function on one surface can be 
transmitted to that of the opposite surface via the (gapless) side surface states. 
Then, the finite-size energy gap decays only algebraically as a function of the thickness. 
The spin Berry phase replaces the periodic boundary condition applied to the electronic 
motion around the cylinder by an anti-periodic boundary condition, 
leading to half-integer quantization of the orbital angular momentum around the cylinder.
\cite{k2}
Such an energy-gap due to phase coherent motion of an electron around the (cylindrical) 
surface is sensitive to introduction of a $\pi$-flux tube piercing the nanowire. 
The sensitivity to $\pi$-flux is a fingerprint of the existence of spin Berry phase, 
which might be directly triggered experimentally in an Aharonov-Bohm type measurement 
recently performed in a system of topological insulator nanowire.
\cite{AB_exp}

As mentioned earlier, the Fermi arc states resemble 
the edge states of a zigzag graphene nano-ribbon
from the viewpoint of topological classification.
But still, as we demonstrate in this paper,
the Fermi arc states exhibit, unlike a 2D graphene layer,
specific spin Berry phase.
In this regard the Fermi arc states show a stronger resemblance
to the helical surface states of a 3D topological insulator.
This paper reveals smoking-gun features of such spin Berry phase in
the Fermi arc states.
We first demonstrate in Sec. II that in a nanowire geometry
electrons in the Fermi arc surface states show
multiple subbands of a partially flat dispersion (see FIG. 1), but 
they are susceptible to a finite-size energy gap
associated with the spin Berry phase.
We then confirm in Sec. III the existence of this spin Berry phase
in the surface effective Hamiltonian by deriving it from
the bulk effective Hamiltonian.
In Secs. IV and V we analyze the system's response to an Aharonov-Bohm flux
as well as to introduction of a screw dislocation,
confirming the existence of spin Berry phase.
We set $\hbar =1$, unless otherwise mentioned.

\begin{figure}
\begin{center}
\includegraphics[width=8cm]{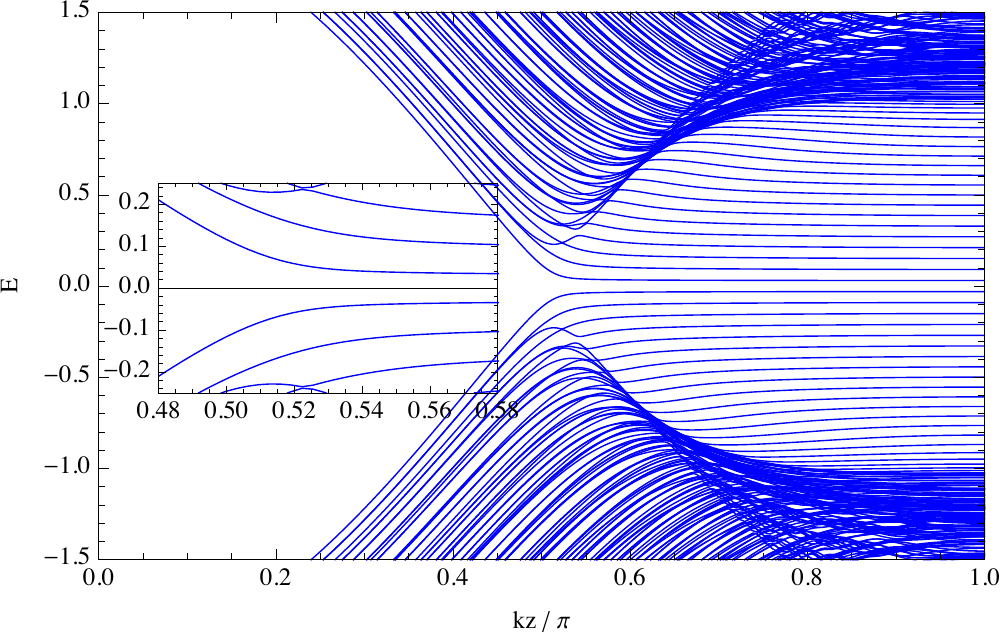}
\end{center}
\caption{Multiple subbands, 
originating from a pair of Fermi arc states 
connecting the two Weyl points ($k_z = \pm k_0 = \pm \pi/2$) 
via the Brillouin zone boundary.
The subbands have a flat band dispersion,
filling the bulk energy gap at $k_z > k_0$.
$A=B=t_z=1$. 
The simulation is done for a system of
square lattice of size: $(N_x, N_y) = (24, 24)$,
whereas periodic in the $z$-direction.
The inset shows details of the spectrum around
$k_z = k_0$ and $E=0$.}
\label{subbands}
\end{figure}

\section{Multiple flat subbands}

Let us consider a simple model of a Weyl semimetal
with a single pair of Weyl nodes on the $k_z$-axis:
\cite{Ran, Balents}
\begin{equation}
H = A (k_x \sigma_x + k_y \sigma_y)
+ M(\bm k) \sigma_z,
\label{H_Weyl}
\end{equation}
where $\bm k =(k_x, k_y, k_z)$, 
\begin{equation}
M(\bm k) = \Delta (k_z) + B (k_x^2 + k_y^2),
\label{mass}
\end{equation}
and we choose
\begin{equation}
\Delta (k_z) = 2 t_z (\cos k_z - \cos k_0).
\label{Delta}
\end{equation}
This is a long-wavelength effective Hamiltonian regarding
the motion in the $x$- and $y$-directions, whereas it can be regarded, 
as for the motion in the $z$-direction, a tight-binding Hamiltonian.
Or, by making the replacements:
\begin{eqnarray}
k_x &\rightarrow& \sin k_x,
\nonumber \\
k_y &\rightarrow& \sin k_y,
\nonumber \\
k_x^2 &\rightarrow& 2(1-\cos k_x),
\nonumber \\
k_y^2 &\rightarrow& 2(1-\cos k_y),
\label{replace}
\end{eqnarray}
the same model can be viewed as a 3D square-lattice tight-binding model.
We will employ this square-lattice implementation for numerical simulations.

The energy spectrum of this effective Weyl model is
characterized by a pair of Weyl points at 
$\bm k = (0, 0, \pm k_0$),
exhibiting a conic dispersion around them.
Besides,
a pair of Fermi arc states appear,
\cite{Ashvin, Balents_kiss, YBKim, Weyl_semi_2, Ran, Balents}
when we put this system into a slab,
say, bounded by the two surfaces, one at $x=0$ and the other as $x=L_x$,
parallel to the $z$-axis.
Let us fix the parameters such that $t_z>0$, $B>0$.
Then for $- k_0 < k_z < k_0$, $\Delta (k_z) > 0$, i.e.,  $\Delta (k_z)/B > 0$.
This means that 
a cross section of Eq. (\ref{H_Weyl}) at a fixed $k_z$ 
in the above range,
describes a trivial 2D band insulator.
Whereas, for $k_0 < k_z < \pi$ and $-\pi < k_z < -k_0$, $\Delta (k_z) < 0$, i.e.,  $\Delta (k_z)/B < 0$.
Then, 
a similar cross section of Eq. (\ref{H_Weyl})
at $k_z$ in one of these ranges,
describes 
a topological (quantized anomalous Hall) insulator with a chiral edge mode.
There appears one single chiral branch
on the $x=0$ side, and another on the $x=L_x$ side,
propagating in the opposite directions;
$-\hat{\bm y}$ and $+\hat{\bm y}$, respectively.
These chiral modes show a linear dispersion, therefore,
of opposite sign,
and cross at $k_y =0$ and at $E=0$,
forming a ``X-shaped'' dispersion $E=E(k_y)$.
If one allows $k_z$ to vary continuously,
then the locus of such an X-shaped dispersion 
forms two planar membranes in the $(k_y, k_z, E)$-space,
always crossing at $k_y =0$ (on the $k_z$-axis) and at $E=0$.
The locus of the crossing point is the Fermi arc,
connecting the two Weyl points $k_z = k_0$ and $k_z = - k_0$ 
via the zone boundary.
Both ends of the two planar membranes are closed by half-conic
structures which appear ``beyond'' the Weyl points;
$k_z < k_0$ and $- k_0 < k_z$.
The entire manifold thus formed envelops the two Weyl cone regions.

Let us then further restrict the system into a nanowire geometry;
the system is restricted not only between $x=0$ and $x=L_x$
but also between $y=0$ and $y=L_y$.
The energy spectrum $E=E(k_z)$ of such a Weyl semi-metallic nanowire is
shown in FIG. \ref{subbands}.
The spectrum shows a series of flat subbands:
\begin{equation}
E(k_z) = E_{\pm 1}, E_{\pm 2}, E_{\pm 3}, \cdots,
\label{E_sub}
\end{equation}
which are remnant of the two planar regions of the Fermi arc manifold.
The flatness of the subbands stems from the fact that
the membrane state has no dispersion in the $k_z$-direction.
These multiple subbands form {\it circular chiral modes},
carrying a spontaneous persistent current around the surface of the wire.
Note that such circular chiral modes do not appear
if one considers a wire perpendicular to the $z$-axis.
In that case 
Fermi arc type surface states do appear on the surface
parallel to the $z$-axis, but they disappear on the side
normal to the $z$-axis. Therefore, the surface states cannot
completely wrap the wire. 
This is quite contrasting to the topological insulator surface states.
The latter, protected by the ``strong'' bulk/edge correspondence, 
appear, irrespectively of the shape and direction of the surface,
and consequently cover the entire surface.
Here, the Fermi arc subbands are indeed in one-to-one correspondence
with the structure of Weyl points in the bulk spectrum.
This correspondence is, however, ``weak'' in the sense that
it depends on the direction of the surface.
As we mentioned earlier,
both the existence and flatness of the Fermi arc subbands are also
topologically protected.

\begin{figure}
\begin{center}
\includegraphics[width=8cm]{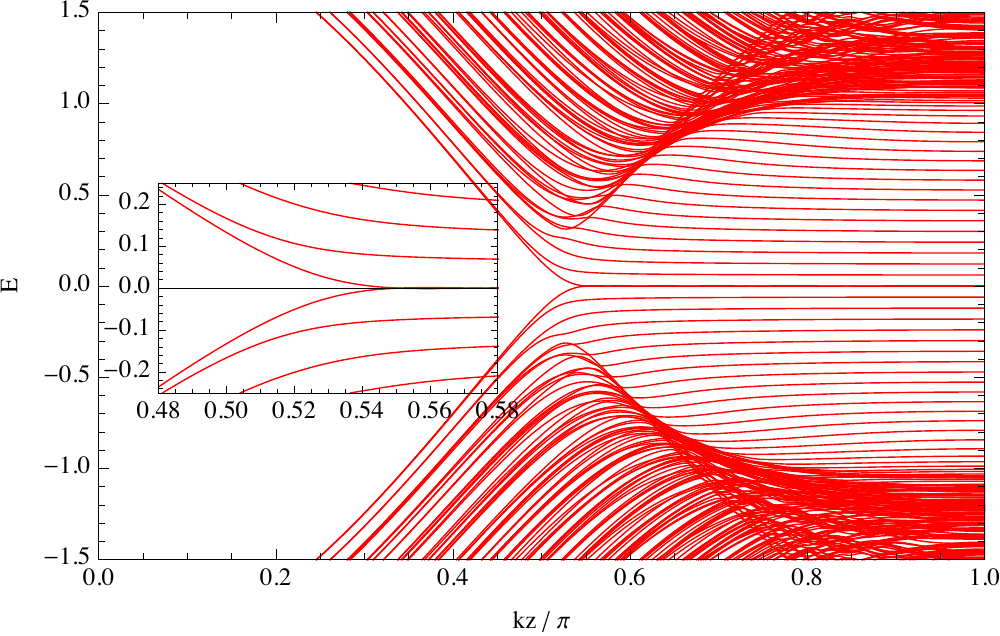}
\includegraphics[width=8cm]{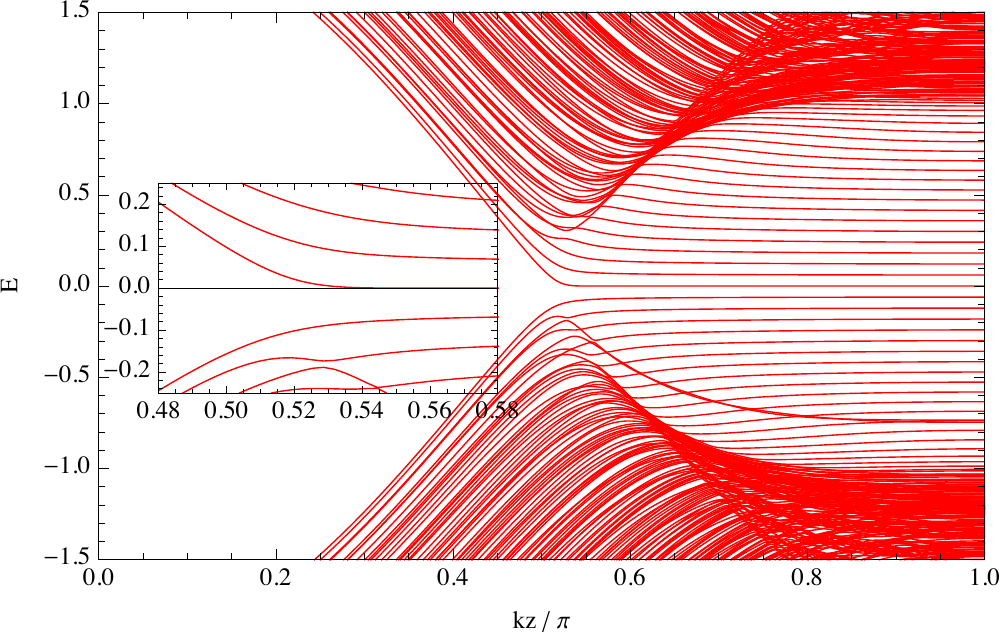}
\end{center}
\caption{Behavior of multiple flat subbands in the presence 
of a flux tube $\Phi = \Phi_0 /2 = hc/(2e)$ 
[$\Phi_0 = hc /e$]
penetrating the nanowire.
Upper: the total flux tube $\Phi = \Phi_0 /2$
pierces a single plaquette at each layer.
Lower: the flux tube is split into two,
each carrying $\Phi /2 = \Phi_0 /4$ in order to avoid a zero-energy
state bound to the flux tube.
$N_x = N_y =24$.}
\label{sub_flux}
\end{figure}

What might be counter-intuitive in FIG. \ref{subbands} is
that the lowest conduction and the highest valence subbands,
$E_{\pm 1} (k_z)$ are
still separated by a finite size energy gap (see its inset).
We see later (FIG. \ref{sub_flux}, upper panel) that one can actually close this gap 
by introducing a flux $\pi$
penetrating the cylinder.
In the light of our knowledge on the helical surface states of a 3D
topological insulator, 
\cite{spin_Berry_1, spin_Berry_2, Mirlin, JMoore, topo_weak, k2}
such a behavior may be naturally attributed to
the existence of spin Berry phase.

Let us come back to FIG. \ref{subbands}, and look into
the subband spectra.
On the $k_z < k_0$ side, the behavior of $E (k_z)$ 
directly results from the bulk spectrum.
Indeed, for $\Delta (k_z)/B > 0$, only bulk solutions are possible.
The ``bulk'' solutions are the solutions of
Eqs. (\ref{H_Weyl}), (\ref{mass}) and (\ref{Delta})
satisfying the boundary condition:
\begin{equation}
\psi (0,y) = \psi (L_x,y) = \psi (x,0) = \psi (x, L_y) =
\left[
\begin{array}{c}
0 \\ 0
\end{array}
\right].
\label{bc}
\end{equation}
The ``plane-wave'' solutions of
Eqs. (\ref{H_Weyl}), (\ref{mass}) and (\ref{Delta})
are
\begin{equation}
|\bm k \pm\rangle_{\rm plane} = e^{i (k_x x + k_y y)} |k_z \rangle |\bm d(\bm k) \pm\rangle,
\label{sol_plane}
\end{equation}
where $\bm k = (k_x, k_y, k_z)$,
and $|k_z \rangle$ is a Bloch state with a crystal momentum $k_z$ in the
$z$-direction. 
$|\bm d(\bm k) \pm\rangle$ represents a spin state
pointing in the direction of
$\bm d (\bm k) = (A k_x, A k_y, M(\bm k))$,
either parallel or anti-parallel,
depending on the index $\pm$.
The plane-wave solution (\ref{sol_plane}) has an energy eigenvalue,
\begin{equation}
E (\bm k) = \pm \sqrt{A^2 (k_x^2 + k_y^2) + M(\bm k)^2}.
\end{equation}
For large enough $\Delta (k_z)$ such that
only the third component of $\bm d (\bm k)$ dominates,
i.e., $\bm d (\bm k) \parallel \hat{\bm z}$,
and one can satisfy the boundary condition Eq. (\ref{bc})
by a simple superposition of $|\bm k \pm\rangle$
with $\bm k = (\pm k_x, \pm k_y, k_z)$, i.e.,
\begin{equation}
|\bm k \pm\rangle_{\rm bulk} \simeq
\sin (k_x x) \sin (k_y y) |k_z \rangle |\hat{\bm z} \pm\rangle,
\label{sol_bulk}
\end{equation}
where $\bm k = (n_x \pi /L_x, n_y \pi /L_y, k_z)$
with $n_x, n_y = 1, 2, \cdots$.
Lowest bulk subbands correspond to
$(n_x, n_y) = (1,1)$, 
$(n_x, n_y) = (1,2)$ and $(2,1)$,
$(n_x, n_y) = (2,2)$,
$(n_x, n_y) = (1,3)$ and $(3,1)$, etc.
In the crossover regime $k_z \sim k_0$, this simple picture
is no longer valid,
but the subbands may be still classified by these quantum numbers.

On the $k_z > k_0$ side, the Fermi arc subbands (\ref{E_sub})
appear in addition to these bulk solutions.
The lowest energy subband solutions in the bulk
merge into the Fermi arc (surface) subbands
in the crossover regime.
The Fermi arc subband solutions
are solutions of 
Eqs. (\ref{H_Weyl}), (\ref{mass}) and (\ref{Delta}) with (\ref{bc})
with $k_x$ and $k_y$ being a complex number.
Their wave functions are localized on the surface of the wire.
Last but not the least, the Fermi arc subbands (\ref{E_sub})
show a flat spectrum,
which appear below the bulk solutions;
$|E|<|\Delta (k_z)|$, 
and exist only in the regime: $\Delta (k_z)/B < 0$.

In the next section, we investigate the nature of such 
surface Fermi arc solutions.
We derive a low-energy effective Hamiltonian which involves only the surface states.
It will become clear that 
the Fermi arc solutions indeed emerge from the bulk effective Hamiltonian,
Eqs. (\ref{H_Weyl}), (\ref{mass}) and (\ref{Delta}),
but importantly, with the spin Berry phase,
which explains the finite size energy gap which we have seen in FIG. \ref{subbands}.

\section{Derivation of the spin Berry phase --- the surface effective Hamiltonian}

To clarify the nature of spin Berry phase, we consider here
a cylindrical nanowire of  radius $R$ extended along the $z$-axis:
$x^2 + y^2 \le R$.
We start from the same bulk effective Hamiltonian,
Eqs. (\ref{H_Weyl}), (\ref{mass}) and (\ref{Delta}),
but in order to extract relevant information on the surface electronic states
we divide it into two components,
\cite{k2, H_surf_1, H_surf_2, Shen_NJP, Shen_PRL, PRB2010}
$H = H_\perp + H_\parallel$,
where
$H_\perp$ ($H_\parallel$) describes electronic motion
perpendicular (tangential) to the cylindrical surface.
Eqs. (\ref{H_Weyl}), (\ref{mass}) and (\ref{Delta}) represent
an effective theory for $k_x, k_y \ll 1$, but 
there is no restriction on $k_z$.
Here, we consider the case: $k_z > k_0$ such that $\Delta (k_z)/B < 0$,
and expand it as $k_z = k_z^{(0)} + p_z$.
Introducing,
\begin{equation}
k_r = -i {\partial \over\partial r}, \ \ 
k_\phi = -i {1\over r} {\partial \over\partial \phi},
\end{equation}
conjugate to the cylindrical coordinates:
$r = \sqrt{x^2 + y^2}$,
$\phi = \arctan {y\over x}$,
one can express $H_\perp$ and $H_\parallel$ as
\begin{eqnarray}
H_\perp &=& H_\perp (k_r) = H |_{k_\phi=0, k_z = k_z^{(0)}},
\nonumber \\
H_\parallel &=& H_\parallel (k_\phi, p_z).
\end{eqnarray}

In order to derive the surface effective Hamiltonian,
we first have to construct a base solution,
the Fermi arc solution in the present case, 
satisfying the given boundary condition,
\begin{equation}
|\psi (r=R, \phi, z) \rangle =
\left[
\begin{array}{c}
0 \\ 0
\end{array}
\right].
\label{bc_R}
\end{equation}
Such a base solution is found by solving the electron dynamics
perpendicular to the surface,
\begin{equation}
H_\perp |\psi_\perp \rangle = E_\perp |\psi_\perp \rangle,
\label{perp_eigen}
\end{equation}
where
$H_\perp$ reads explicitly
\begin{equation}
H_\perp = 
\left[
\begin{array}{cc}
M_\perp & A k_r e^{-i\phi}\\ 
A k_r e^{i\phi} & - M_\perp
\end{array}
\right].
\end{equation}
Here, we have decomposed the mass term into
\begin{eqnarray}
M_\perp &\simeq& \Delta (k_z^{(0)}) + B k_r^2,
\nonumber \\
M_\parallel &=& B k_\phi^2 - 2 t_z \sin k_z^{(0)} p_z.
\end{eqnarray}
The Laplacian in the cylindrical coordinates
has another contribution, $(1/r)\partial / \partial r$.
Here, we neglect this first-order derivative term, 
keeping only the term $\partial^2 / \partial r^2$.
\cite{k2}
This is justified when the radius $R$ of the cylinder is
sufficiently larger than the penetration depth [$\kappa^{-1}$, see Eq. (\ref{kappa})] 
of the surface state.

We search for solutions of Eq. (\ref{perp_eigen}),
which has an energy $E_\perp$, in the range:
$- \Delta (k_z) < E_\perp < \Delta (k_z)$,
and takes the following form,
\begin{equation}
\label{kappa}
|\psi_\perp \rangle \simeq e^{\kappa (r-R)} | E_\perp, \kappa \rangle,
\end{equation}
where
\begin{equation}
| E_\perp, \kappa \rangle =
\left[
\begin{array}{c}
E_\perp + M_\perp (\kappa)\\ 
-i \kappa A e^{i\phi}
\end{array}
\right],
\label{spinor}
\end{equation}
and $\kappa>0$.
For a given energy $E_\perp$, $\kappa$ has two positive solutions,
$\kappa = \kappa_\pm$, satisfying
\begin{eqnarray}
\label{energy}
E_\perp^2 &=& M_\perp (\kappa)^2 - A^2 \kappa^2,
\\
M_\perp (\kappa) &=& \Delta - B \kappa^2.
\nonumber 
\end{eqnarray}
Composing a linear combination of
these two base solutions, one can construct a hypothetical
wave function,
\begin{equation}
|\psi_\perp \rangle =
c_1 e^{\kappa_+ (r-R)} |E_\perp, \kappa_+ \rangle +
c_2 e^{\kappa_- (r-R)} |E_\perp, \kappa_- \rangle,
\label{sol_hypo}
\end{equation}
which should be matched with the boundary condition (\ref{bc_R}),
i.e.,
\begin{equation}
\det
\left[
\begin{array}{cc}
E_\perp + M_\perp (\kappa_+) & E_\perp + M_\perp (\kappa_-)
\\
- i\kappa_+ A e^{i\phi} & - i\kappa_- A e^{i\phi}
\end{array}
\right]=0.
\label{det}
\end{equation}
Note that the two wave functions 
$|\psi  \rangle$ and  $|\psi_\perp \rangle$ are related
by $|\psi  \rangle =  \psi_\parallel (\phi, z) |\psi_\perp \rangle$.
Since $\kappa_+ \neq \kappa_-$, the condition (\ref{det}) 
simplifies (after some algebra) to
\begin{equation}
E_\perp (\Delta + E_\perp) =0.
\label{E=0}
\end{equation}
Recall that at fixed $k_z$ the Fermi arc solution appears in the bulk gap:
$- \Delta (k_z) < E_\perp < \Delta (k_z)$, i.e., 
Eq. (\ref{E=0}) imposes $E_\perp = 0$.
Substituting this back to Eq. (\ref{energy}), one finds
$M_\perp = \pm \kappa A$.
For $E_\perp = 0$ and $M_\perp = \kappa A$, Eq. (\ref{spinor}) becomes
\begin{equation}
| E_\perp =0, \kappa \rangle = M_\perp
\left[
\begin{array}{c}
1\\ 
-i e^{i\phi}
\end{array}
\right],
\label{spinor_1}
\end{equation}
and the two solutions for $\kappa$ becomes
\begin{equation}
\kappa_\pm = {A \pm \sqrt{A^2 + 4B\Delta}\over 2B},
\end{equation}
which is concistent with the condition: $\kappa_\pm >0$
(recall that $B\Delta < 0$).
The other choice, $M_\perp = -\kappa A$
is not compatible with this requirement.

Thus the normalized Fermi arc base solution is found to be
\begin{equation}
|\psi_\perp \rangle =
\rho (r)
\left( 
e^{\kappa_+ (r-R)} - e^{\kappa_- (r-R)}
\right)
\left[
\begin{array}{c}
1\\ 
-i e^{i\phi}
\end{array}
\right],
\label{sol_base}
\end{equation}
where 
\begin{equation}
\rho (r) \simeq
\sqrt{\kappa_+\kappa_-(\kappa_+ +\kappa_-) \over 2\pi R}
{e^{\kappa_+ (r-R)} - e^{\kappa_- (r-R)}
\over |\kappa_+ - \kappa_-|}
\label{rho_r}
\end{equation}
($\kappa_\pm R \gg 1$ assumed).
Eq. (\ref{sol_base}) is a remarkable result, indicating that the surface spin state is
\begin{equation}
|\hat{\bm \phi} - \rangle
=
{1\over \sqrt{2}}
\left[
\begin{array}{c}
e^{-i\phi/2} \\ 
-i e^{i\phi/2}
\end{array}
\right],
\label{spin_surf}
\end{equation}
diagonalizing a spin operator in the direction of $\hat{\bm \phi}$
with an eigenvalue $-\hbar /2$,
where $\hat{\bm \phi}$ is
a unit vector pointing to the azimuthal direction,
\begin{equation}
\hat{\bm \phi}=
\left[
\begin{array}{c}
- \sin \phi \\ 
\cos \phi
\end{array}
\right].
\end{equation}
The electronic spin in the Fermi arc state
is locked in the direction (anti-) parallel to that of $\hat{\bm \phi}$,
and when an electron goes around the cylinder in the anti-clockwise
direction,
it also rotates, following the curved surface of the cylinder,
locked in the direction of $- \hat{\bm \phi}$.
After a complete $2\pi$ rotation,
the electron goes back to the original position on the
cylinder,
the spin also comes back its original state, but with an additional phase of $\pi$.
This may not be clear from Eq. (\ref{sol_base}),
since it is written in the single-valued representation.
\cite{k2}
Yet, information on the double-valuedness of spin is safely encoded
in the surface effective Hamiltonian in the form of spin Berry phase, as we see below.
Notice also that here,
in contrast to the case of helical surface states of a topological insulator,
\cite{spin_helical, Hasan_Kane}
the surface spin state is not helical.
It is rather ``chiral'', pointing to the azimuthal direction
of the cylinder independently of the value of $k_z$.
The spin direction is locked indeed anti-parallel to the group velocity of
the surface mode.
Let us finally see such {\it chiral} spin-to-surface locking leads, indeed,
to the appearance of spin Berry phase.
Using the base solution (\ref{sol_base}), we calculate the expectation value of
\begin{equation}
H_\parallel = 
\left[
\begin{array}{cc}
M_\parallel &  -i A e^{-i\phi} k_\phi
\\ 
-i A e^{i\phi} k_\phi & - M_\parallel
\end{array}
\right],
\end{equation}
to find
\begin{equation}
H_{\rm surf} =
\langle \psi_\perp| H_\parallel |\psi_\perp \rangle 
= {A\over R}\left( -i {\partial \over\partial \phi} + {1\over 2} \right).
\label{H_surf}
\end{equation}
The low-energy electron dynamics along the surface is thus determined by
the eigenvalue equation,
\begin{equation}
H_{\rm surf} \psi_\parallel (\phi, z) = E_\parallel \psi_\parallel (\phi, z)
\label{surf_eigen}
\end{equation}
where
$\psi_\parallel (\phi, z) = e^{i k_\phi \phi} e^{i k_z z}$.
The periodic boundary condition around the wire,
\begin{equation}
\psi_\parallel (\phi +2\pi, z) = \psi_\parallel (\phi, z),
\label{pbc}
\end{equation}
requires that $k_\phi$ be an integer.
The spin Berry phase term, i.e., factor $1/2$ in Eq. (\ref{H_surf})
plays, then the role of shifting the surface electron spectrum by 
a half of the finite-size energy gap,
\begin{equation}
E_{\rm surf} = {A\over R}\left( k_\phi + {1\over 2} \right).
\label{E_surf}
\end{equation}

Recall that
the origin of this $1/2$ spin Berry phase term is that
the spin in the Fermi arc surface state is locked in the direction of
Eq. (\ref{spin_surf}).
This is quite contrasting to the spin state of the bulk solution;
see Eqs. (\ref{sol_plane}) and (\ref{sol_bulk}).
In the crossover regime $k_z \sim k_0$,
the bulk spin state $|\bm d \pm\rangle$
on the $k_z < k_0$ side
evolves into the locked surface spin state,
Eq. (\ref{spin_surf}).
In parallel with this evolution in spin space,
the wave function of lowest energy bulk subbands 
tend to become localized around the boundary.
In spectrum, they merge into the Fermi arc (surface) subbands.

\begin{figure}
\begin{center}
\includegraphics[width=8cm]{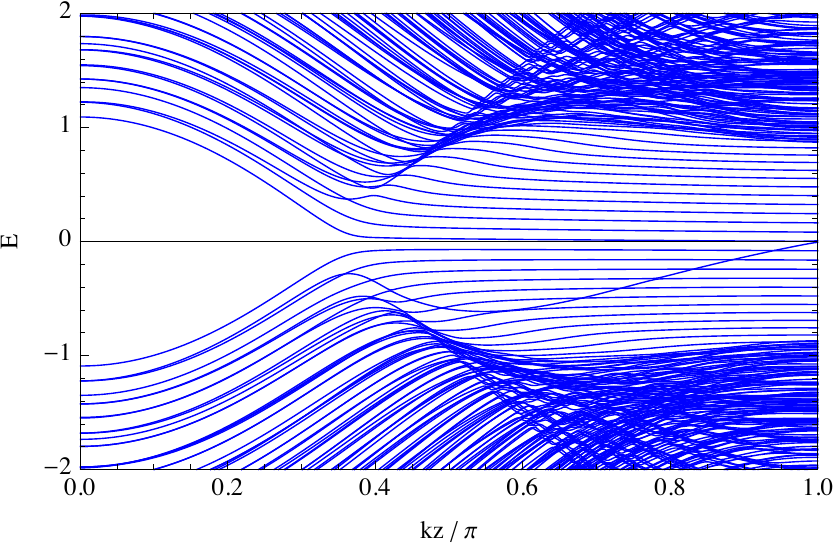}
\includegraphics[width=8cm]{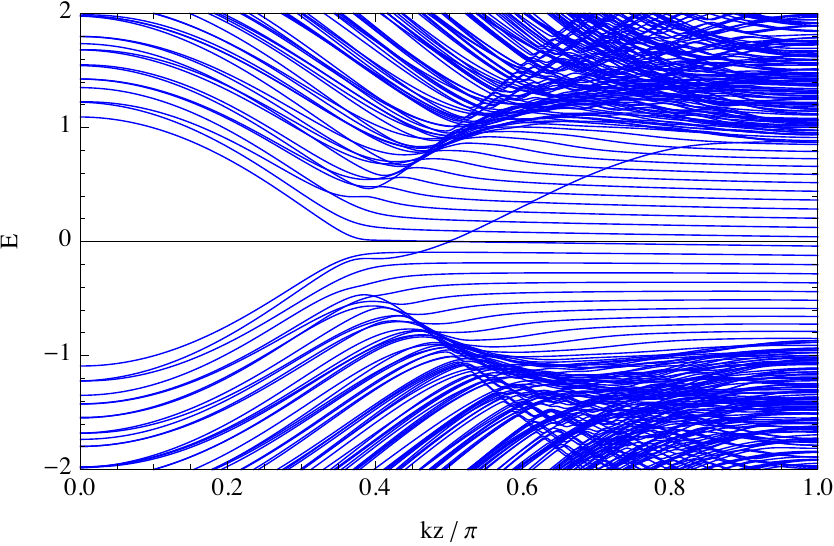}
\includegraphics[width=8cm]{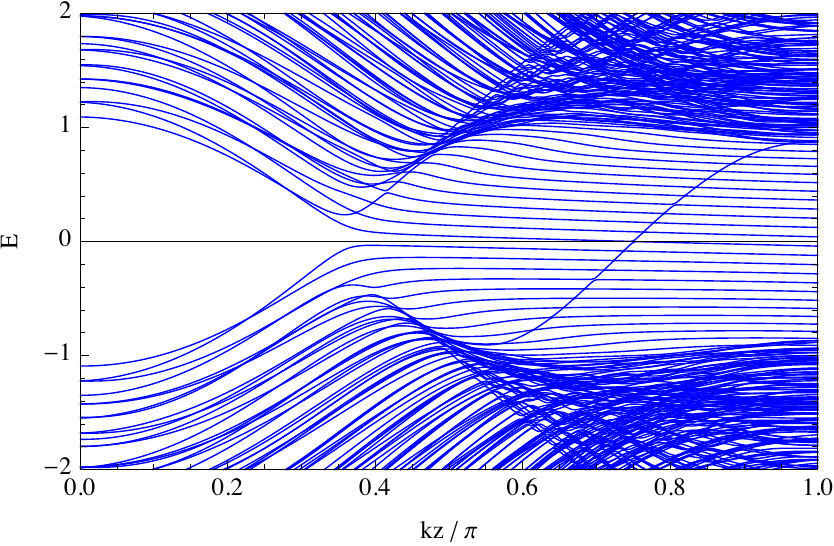}
\end{center}
\caption{Multiple subbands deformed by a crystal dislocation
with a Burgers vector $\bm b=(0,0,b)$, 
where $b=1$ for the top, $b=2$ for the central, and
$b=4$ for the bottom panel. $k_0 = \pi/3$,
$N_x = N_y =16$.}
\label{sub_screw}
\end{figure}

\section{Numerical confirmation of the spin Berry phase}

Let us verify the existence of spin Berry phase in numerical experiments.
We reconsider the nanowire geometry with a square cross section,
which has already appeared in Sec. II.
Here, to confirm the existence of spin Berry phase, we
introduce a flux tube $\Phi$ piercing the nanowire.
In FIG. $\ref{sub_flux}$
the energy spectrum in the presence of an infinitely thin 
$\pi$-flux tube,
carrying a magnetic flux $\Phi=\Phi_0 /2 = hc /(2e)$,
with $\Phi_0 = hc /e$ being the unit flux quantum,
is shown, to be compared with
the spectrum in the absence of flux (FIG. \ref{subbands}).
The upper panel of FIG. $\ref{sub_flux}$
shows the spectrum 
when the total flux $\Phi_0 /2 = hc /(2e)$ pierces,
at each cross section, a {\it single} plaquette.
The obtained spectrum shows a gapless dispersion
with doubly degenerate zero-energy states.
This behavior is indeed quite contrasting to the gapped
spectrum of FIG. $\ref{subbands}$,
which is reminiscent of an analogous behavior 
due to the effect of spin Berry phase 
on the surface of a topological insulator.
\cite{topo_weak}
Still it is not simply explained by the effective surface theory,
Eqs. (\ref{H_surf}), (\ref{surf_eigen}), (\ref{pbc}) and (\ref{E_surf}),
which we have derived in the last section.
Notice that 
in the presence of a flux $\Phi_0 /2 $ penetrating the cylinder,
the surface effective Hamiltonian (\ref{H_surf}) is replaced by,
\begin{equation}
H_{\rm surf} 
= {A\over R}\left( -i {\partial \over\partial \phi} + {1\over 2} - {\Phi \over \Phi_0} \right),
\label{H_surf_2}
\end{equation}
resulting in a shift of the spectrum,
\begin{equation}
E_{\rm surf} (\Phi)= {A\over R}\left( k_\phi + {1\over 2} - {\Phi \over \Phi_0} \right).
\label{E_phi}
\end{equation}
Thus, according to the effective surface electron dynamics,
Eqs. (\ref{H_surf_2}), (\ref{surf_eigen}), (\ref{pbc}) and (\ref{E_phi}),
the subband spectra are {\it uniformly} shifted
both for $k_\phi$ positive and negative (or null) integer.

Then, where does the degenerate $E=0$ pair come from?
The answer to this question is almost obvious if one looks into
spatial distribution of the corresponding wave function.
As a general consequence of the $1/2$ spin Berry phase term,
a series of plaquettes (aligned in the $z$ direction) 
each penetrated by a $\pi$-flux tube
always hosts a zero-energy bound state.
(A possibly related effect in a strong topological insulator
can be found in Ref. \cite{Franz}.)
The existence of such a bound state might be clear
from Eq. (\ref{E_phi}).
As a crude approximation, 
one can regard the series of plaquettes which accommodates the 
$\pi$-flux as 
a cylinder of a radius $r_0 \sim 1$ (the lattice constant).
Taking into account (though this is irrelevant to the discussion here)
that the surface state localized around this
cylinder will have an opposite chirality (propagating direction),
the effective surface Hamiltonian (\ref{H_surf}),
may be modified to describe such a bound state as
\begin{equation}
H_{\rm bound} = 
- {A\over r_0}\left( -i {\partial \over\partial \phi} + {1\over 2} - {\Phi \over \Phi_0} \right).
\label{H_bound}
\end{equation}
In Eqs. (\ref{H_surf_2}), (\ref{H_bound})
the electrons feel the same flux, only the propagating direction is opposite.
We may write the corresponding wave function as
$\psi_\parallel (\phi, z) = e^{i n \phi} e^{i k_z z}$,
with a quantum number $n$ associated with the orbital motion
around the flux tube, rather than $k_\phi$ 
to make a distinction between the two.
Then, Eq. (\ref{E_phi}) becomes,
\begin{equation}
E_{\rm bound} (\Phi) = - {A\over r_0}\left( n + {1\over 2} - {\Phi \over \Phi_0} \right).
\label{E_bound}
\end{equation}
In any case, cancellation of the $1/2$ spin Berry phase term by the $\pi$-flux
implies the existence of a zero-energy bound state.
Of course, since $r_0 \sim 1 \ll R$,
in the spectrum of Eq. (\ref{E_bound}) 
only the $E=0$ ($n = 0$) state is relevant 
in the energy scale $A/R$ of the finite-size energy gap; {\it cf.} Eq. (\ref{E_phi}),
and appears in the window of bulk energy gap.
Such a bound state along the flux tube is degenerate with
the $E=0$ subband state of Eq. (\ref{E_surf}) with $k_\phi = 0$
and explain the two-fold degeneracy of $E=0$ state in FIG. \ref{sub_flux}. 

However, 
if one's purpose is to see simply the effects of spin Berry phase,
one can avoid this complexity.
The lower panel of FIG. $\ref{sub_flux}$
shows a spectrum when the system is always
pierced by a $\pi$-flux tube, with a magnetic flux
$\Phi = \Phi_0 /2 = hc /(2e)$,
but {\it split into two};
each of the half flux $\Phi /2 = \Phi_0 /4 = hc /(4e)$ 
pierces a different plaquette.
One can still assume a bound state along such
a half flux tube, and estimate its energy.
For a cylinder penetrated by a half flux tube, Eq. (\ref{E_bound})
modifies to,
\begin{equation}
E_{\rm bound} (\Phi) \rightarrow E_{\rm bound} (\Phi /2)
=  - {A\over r_0}\left( n + {1\over 2} - {\Phi/2 \over \Phi_0} \right).
\end{equation}
Clearly,
for $\Phi=\Phi_0 /2$ and $r_0 \sim 1 \ll R$,
there exists no bound state in the scale of finite-size energy gap $A/R$.
The $n=0$ bound state is sent to the high-energy spectrum
(one can actually see this in the lower panel of FIG. \ref{sub_flux}).
Thus, the low-energy (at the energy scale of $A/R$) spectrum of such a system is simply determined by
Eq. (\ref{E_phi}).
The obtained data depicted in the lower panel of FIG. \ref{sub_flux}
indeed show such a behavior consistent with Eq. (\ref{E_phi}),
with a single subband state 
located precisely at $E=0$ (see the inset).

An alternative way to verify the existence of spin Berry phase
is to see system's response to crystal deformation;
introduction of a screw dislocation.
A screw dislocation plays, fundamentally, 
a role similar to the magnetic flux
we have considered above,
\cite{Ran_nphys, topo_weak}
but its effect on the subband spectrum is
superficially much different, 
as shown in FIG. \ref{sub_screw}.
Suppose that 
the underlying crystal is deformed by
a screw dislocation along the axis of the wire ($z$-axis);
its Burgers vector is $\bm b=(0,0,b)$
($b=\pm 1, \pm 2, \cdots$).
As opposed to a magnetic flux which twists
the phase of an electronic wave function uniformly,
a crystal dislocation introduces a phase shift
which depends on the crystal momentum $k_z$,
in the direction of the Burgers vector,
\begin{equation}
\psi_\parallel (\phi, z) = e^{i (k_\phi - k_z b/ (2\pi)) \phi} e^{i k_z z},
\label{psi_b}
\end{equation}
introducing a finite slope into the subband spectrum,
\begin{equation}
E_{\rm surf} (k_z b) = {A\over R}\left( k_\phi + {1\over 2} - {k_z b\over 2\pi} \right).
\label{E_kzb}
\end{equation}
Notice that the $k_\phi$-th subband
intersects with the $E=0$ line at 
\begin{equation}
k_z b = (2 k_\phi + 1) \pi,
\end{equation}
similarly to the $\pi$-flux case, 
but here this occurs only at such discrete values of $k_z$.

Is that all that a dislocation line does to the subband spectrum?
No, of course not.
In each panel of FIG. \ref{sub_screw}
one can recognize an isolated mode
which has a slope opposite to all the other subband states.
This is again due to a bound state formed along a dislocation line.
Similarly to the case of a $\pi$ flux tube 
piercing a single plaquette at each $(x,y)$-layer,
a series of plaquettes 
penetrated by the dislocation line may be
regarded as a cylinder of radius $r_0 \sim 1$.
The spectrum of subband states associated with such a dislocation line
reads
\begin{equation}
E_{\rm bound} (k_z b) = - {A\over r_0}\left( n + {1\over 2} - {k_z b \over 2\pi} \right).
\label{E_disloc}
\end{equation}
Again, for $r_0 \sim 1$
only at most a few subbands, satisfying a zero-energy condition,
\begin{equation}
k_z b = (2 n + 1) \pi,
\end{equation}
are visible in the relatively small window of the bulk spectrum.
Such subbands
has a steep positive slope as a function of $k_z$ opposite to 
all the other subbands described by Eq. (\ref{E_kzb}) since $R \gg 1$;
i.e., the bound state(s) along the dislocation line is (are)
propagating modes.
Notice that 
in the two upper panels of FIG. \ref{sub_screw} (cases of $b=1, 2$)
crossing of the two subbands, (\ref{E_kzb}) and (\ref{E_disloc}),
occurs at $k_\phi = n =0$, whereas 
in the bottom panel ($b=4$ case),
the same crossing which falls on the Fermi arc region
is at $k_\phi = n =1$, indicating that
$k_z = (3/4) \pi$.

We have seen that a  (square cross sectional) cylinder
pierced by a dislocation line should be regarded
as a pair of concentric cylinders with radii, 
$r_0 \sim 1$ and $R \gg 1$.
This allows us to interpret the spectra shown in FIG. \ref{sub_screw}
as a superposition of two contributions; one from the subbands
localized in the vicinity of the outer surface: Eq. (\ref{E_kzb}), 
and the other from a bound state along a dislocation line:
Eq. (\ref{E_disloc}).
To double check the validity of such an interpretation
in terms of the bound state along a dislocation line,
we consider in the next section an extreme example, 
in which only the bound states appear.

\begin{figure}
\begin{center}
\includegraphics[width=8cm]{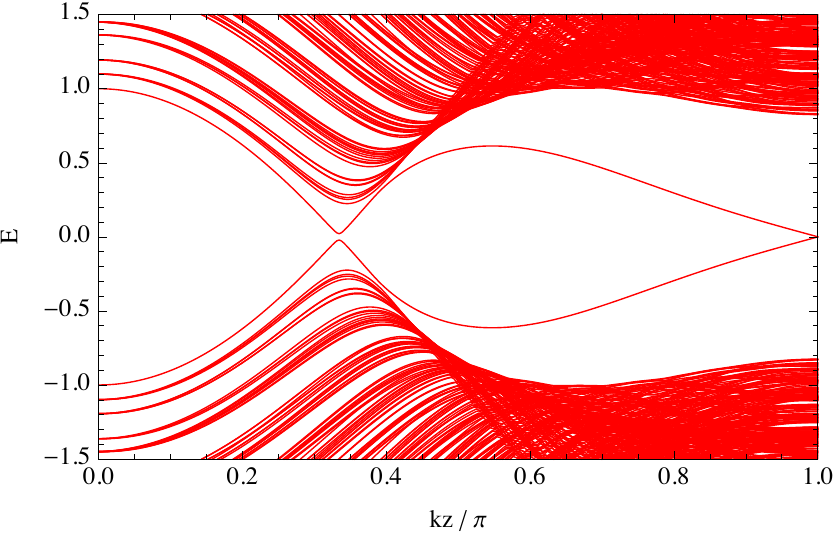}
\includegraphics[width=8cm]{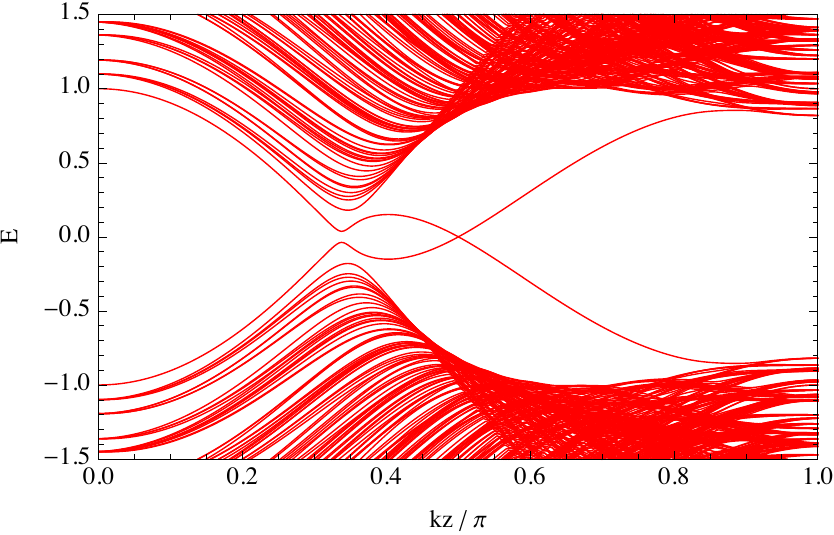}
\includegraphics[width=8cm]{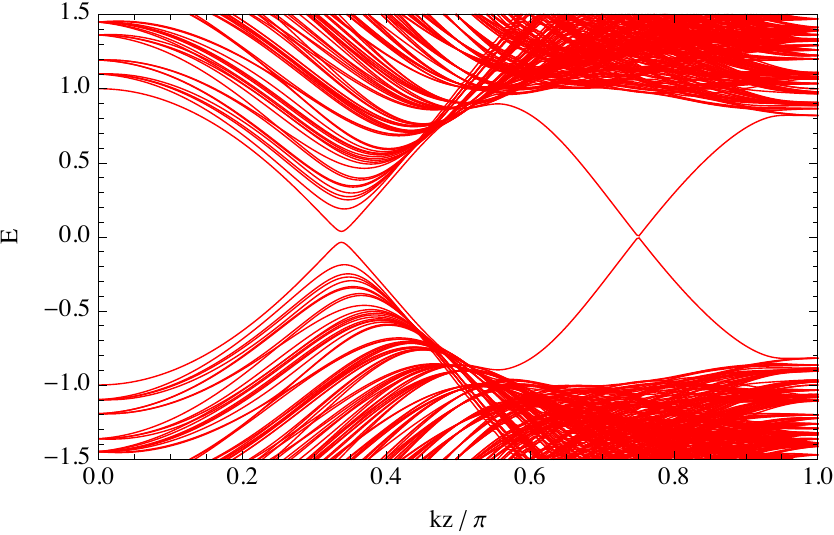}
\end{center}
\caption{1D chiral modes along a pair of dislocation lines.
$b=1$ (top), $b=2$ (central) and $b=4$ (bottom). 
$k_0=\pi/3$, $N_x = N_y =24$.}
\label{screw_torus}
\end{figure}

\section{Protected 1D chiral mode along a dislocation line}

Let us finally consider a slightly different geometry in which 
a pair of screw dislocations, one parallel, the other anti-parallel with
the $z$-axis, penetrate a triply periodic (surfaceless) system;
the two dislocation lines are spatially well-separated.
On a planar region bounded by the two dislocation lines
the crystal is dislocated in the $z$-direction by $b$.
This attributes to each dislocation line
a pair of Burgers vectors, $\bm b = (0,0 , b)$ and $\bm b = (0,0 , -b)$.
As we have already seen, 
such a situation is readily described by the cylinder model
we have considered in Sec. IV.
A minor but not unimportant difference from the previous case is
that here the two cylinders are parallel,
and not concentric.
Therefore, on the surface of the two cylinders the orbital motion of an electron 
around the cylinder is in the same anti-clockwise direction.
The low-energy electron dynamics on the surface of the
two cylinders are described by the same surface effective Hamiltonian (\ref{H_surf}).
Only the Burgers vector differs, and modulates the phase of the
electronic wave function in different ways;
along each of the dislocation line, Eq. (\ref{psi_b}) should be replaced by,
\begin{eqnarray}
\psi_1 (\phi, z) = e^{i (n_1 - k_z b/ (2\pi)) \phi} e^{i k_z z},
\label{psi_1} \\
\psi_2 (\phi, z) = e^{i (n_2 + k_z b/ (2\pi)) \phi} e^{i k_z z}.
\label{psi_2}
\end{eqnarray}
The corresponding bound state spectra read,
\begin{eqnarray}
E_1 (k_z b) &=& {A\over r_0}\left( n_1 + {1\over 2} - {k_z b \over 2\pi} \right),
\label{E_1} \\
E_2 (k_z b) &=& {A\over r_0}\left( n_2 + {1\over 2} + {k_z b \over 2\pi} \right).
\label{E_2}
\end{eqnarray}
Again, since $r_0 \sim 1$
only the lowest energy subbands, satisfying the zero-energy condition,
\begin{equation}
k_z b = (2 n_1 + 1) \pi,\ \
- k_z b = (2 n_2 + 1) \pi,
\label{n1n2}
\end{equation}
are relevant in the spectrum.
Some concrete examples of calculated spectra for such a system
are shown in FIG. \ref{screw_torus}
for $b=1$ (top), $b=2$ (central) and $b=4$ (bottom). 
Each spectrum exhibits a pair of chiral modes,
which are identified as the states represented by
Eqs. (\ref{psi_1}) and (\ref{psi_2}) with
$n_1$ and $n_2$ satisfying Eqs. (\ref{n1n2})
in the Fermi arc region: $k_0 < k_z < \pi$.
For the two upper panels (cases of $b=1, 2$),
$n_1=0$, $n_2 = -1$, i.e.,
the two chiral modes intersect at 
$k_z = \pi /b$ and at $E=0$.
As for the last panel ($b=4$),
$n_1=1$, $n_2 = -2$,
indicating that the intersection occurs at
$k_z = (3/4)\pi$.
Note that this type of gap closing 
always occurs at $E=0$ and at the same $k_z$ points
uniquely determined by the Burgers vector.
Such a feature is 
model independent, and in this
sense these chiral modes are protected.
Notice, in contrast,
that projection of the 3D Weyl point onto the 1D Brillouin zone 
($k_z = \pi /3$ in FIG. \ref{screw_torus})
is gapped by the screw dislocation.

It is also interesting that such a pair of 
zero energy bound states have a dispersion in the $k_z$-direction;
they have a finite group velocity of order $\sim A$, 
and are 
propagating along the dislocation line, but in the opposite directions on each of the
dislocation lines.
In this sense we call each of them a 1D {\it chiral} mode.
Note that here the meaning of ``chiral'' is different from
when we used the same word to describe
the chiral property of the Fermi arc surface state or its subbands.
Indeed, on each of the dislocation line 
the circular orbital motion of an electron in the Fermi arc state
(around a hypothetical cylinder of radius $r_0 \sim 1$)
is in the same anti-clockwise direction,
but following a spiral which evolves in the opposite direction
($+z$ or $-z$) reflecting the opposite direction of the
Burgers vector.

\section{Conclusions}

We have studied electronic states of a 3D Weyl semimetal,
which serve, in this regard,
as the 3D counter part of graphene.
Naturally the corresponding Fermi arc surface states
could be regarded as a 2D version of 1D edge modes
with a flat dispersion,
which are known to exist in the zigzag edge nano-ribbon.

This paper, however, points out a crucial difference between the two systems.
The Fermi arc surface states exhibit a specific type of
(chiral) spin-to-surface locking.
This manifests as spin Berry phase when one considers a curved surface,
e.g., a cylindrical surface.
In 1D edge states of the zigzag nano-ribbon
the edge pseudo-spin state is determined rather by the structure of the edge.
\cite{zigzag_1, zigzag_2}
The spin Berry phase has been regarded as a hallmark property of
the helical surface states of a topological insulator.
In this paper we have demonstrated that
in the case of 3D Weyl semimetal, the existence of
peculiar spin Berry phase in the Fermi arc state leads to
a number of unusual finite size effects:
\begin{enumerate}
\item 
The nanowire spectrum shows a feature of
multiple subbands, which is gapped at $E=0$ (at the level of Weyl points).
\item
In the case of triply periodic surfaceless system,
a protected gapless chiral mode appears along a dislocation line
and dominates the low-energy transport.
\end{enumerate}
As a general remark, we have emphasized that
whenever a system bears a surface state involving a
spin Berry phase,
a series of plaquettes pierced by a $\pi$-flux tube always
hosts a zero-energy bound state.
The second statement above is
a specific version of this general phenomenon
in the case of a 3D Weyl semimetal bearing Fermi arc
states.

\acknowledgments
KI acknowledges Y. B. Kim and A. Schnyder for useful discussions.
The authors are supported by KAKENHI;
KI by the ``Topological Quantum Phenomena'' [No. 23103511],
and YT by a Grant-in-Aid for Scientific Research (C) [No. 21540389].

\bibliography{FA_r2_v2x}

\end{document}